# ORIENTATIONS OF THE GIANT'S CHURCHES IN OSTROBOTHNIA, FINLAND
# I

MARIANNA RIDDERSTAD, University of Helsinki, Finland, and
JARI OKKONEN, University of Oulu, Finland

*Introduction*

The so-called Giant's Churches are Neolithic stone structures, unique to the coastal area between Yli-Ii and Närpiö in Ostrobothnia in the western Finland.[1] They date from 2500-2000 BCE, and are concentrated on the ancient seashore.[2] Most of them were built on islands or drumlins on the coast, but are now situated as far as 30 kilometers inland because of the post-glacial rebound. There are 40 to 50 of them, depending on the definition, which is not clear, as their function is not yet known.

The Giant's Churches are large, the length of the long axis differing from about 60 meters to 12 meters, and most often rectangular. On the other hand, the height of their walls is rather low, from about half a meter to about 2 meters in some cases. Most of them have "gates", which are lowerings in the walls, suitable for entering the structure. Some of them also have so-called "sacristies", which are stone cairns either constructed as parts of the walls or situated immediately outside of them.

The function of the Giant's Churches has been a matter of debate for over a hundred years. They have been seen as dwellings, burial sites, temples, fortresses, natural formations, giant cold stores for seal meat and hunting enclosures.[3] Some of the smallest constructions may have been dwellings, but the largest ones would have been impractical for that purpose. No usual signs of permanent inhabitation have been found inside the structures.

The present paper is the first one in the series of studies, where the possible deliberate orientations of the Giant's Churches to celestial events are examined. The results may help to clarify the functions of the structures. In this paper, the orientations of 23 Giant's Churches are presented.

*Measurements*

The orientations of the structures were measured as follows. (1) In rectangular structures (denoted with Q in Table 1), the directions of the walls were measured to determine the orientation of the long axis of each structure. In one case (Linnakangas in Ruukki), which is egg-shaped, the direction of the greatest length of the structure was used. (2) The directions of the gates and, in three cases, the sacristies, were measured as seen from the centre of the structure. The centre of the structure was determined from the measurements of the surrounding walls. In the structures with double walls (Q2), the centre of the structure was determined using the inner walls, if they were clearly detectable. In two cases, Hangaskangas and Kiviojankangas, where the structures are so-called 'open rectangular' (Open R), i.e., lacking one or two walls to create a full rectangle, the directions of the walls and the directions towards the ends of the walls were measured. In Hangaskangas, the directions towards the ends of the walls were measured as seen from the sacristy in the 'bottom' end of the open structure, whereas in Kiviojankangas, the directions towards the ends of the open walls were measured from the single gate in the long wall.

---

[1] Europaeus 1913, p. 90-91.
[2] See: Okkonen 2003.
[3] See: Okkonen 2003, p. 57, 60, 124, 131-133.

*Orientations and discussion*

The results of the measurements of the orientations of the long axes, gates and sacristies of 23 Giant's Churches are presented in Table 1.

As the Giant's Churches were originally built on the seashore or on islands, the original horizon towards the west was flat in almost all cases. Because they were most often built on the highest points of the natural formations protruding out of the former sea floor also in the east, also the eastern the horizon height was, in most cases, zero. Only in some cases the orientations were observed to be towards a nearby elevated point, another rocky island or drumlin. In these cases, the height of the horizon is given in Table 1. The existence of trees on the islands would, of course, have affected the horizon height, but, at this point of research, this effect is impossible to estimate, as it is not yet known whether many of the structures were built when the locations were still bare outer islands, or already parts of the forest-covered coastal region.

The greatest source of error in the study comes from the level of preservation of the structures, which affects the estimated locations of the centres and gates of the structures. Some of the Giant's Churches, like Kastelli, are rather well-preserved and have been cleared from the surrounding forest. In three cases (marked with * in Table 1), the digging of a sand pit had or forest ploughing had destroyed parts of the structures. In one case (Storbacken), the structure was so thickly covered in moss and lichen that no gates could be reliably observed. In four cases (marked with ? in Table 1), the existence or original direction of a gate or a sacristy, towards which the orientation was measured, is unclear due to level of preservation.

The solar events considered in this study are the solstices, the equinoxes and the so-called Mid-Quarter Days. As the two Mid-Quarter Days surrounding each solstice have the same solar positions, only two of them need to be considered. In Finland, the most important festivals coinciding with these four parts of the year have traditionally been Vappu (St. Valborg's Day) in May and Kekri, which was the ancient festival of the dead predating historical times, celebrated in November.

These eight main solar dates of the year correspond to ten different rising and setting points of the sun on the sky. At the latitude of Raahe, the rising and setting points of the sun at these dates correspond to total of 11% of the width of the horizon. With the +-5 deg error limits, the rising and setting points correspond to 36% of the horizon.

As Table 1 shows, within the error limits of +-5 degrees, twelve of the Giant's Churches, i.e., about half of them, have their long axes oriented to solar events.

Of the 58 orientations towards gates, sacristies and the ends of walls in open structures, 29 are oriented to solar events within the +-5 deg error limits. Within the error limits of +-1 deg, 19 of the orientations measured are to solar events.

Of the 29 solar orientations, 6 are to the sunrise or sunset at the winter solstice, 4 at the summer solstice, 3 at the equinoxes and 16 at Vappu and Kekri.

The orientations of the sacristies must be, at this point, considered secondary in importance to the gate orientations. However, our results indicate that their orientations may be significant and should be further studied in the future.

The northernmost of the sites is Rajakangas in Haukipudas (65.2 deg N, 25.8 deg E), and the southernmost is Storbacken in Evijärvi (63.5 deg N, 23.3 deg E). These latitudes are unique in the Northern Hemisphere, because, between them, the daily path of the sun is very close to the horizon at the winter solstice, and at the summer solstice, the sun barely sets. Between these latitudes, the azimuth of the rising point of the upper limb of the sun changes from 160 to 152 deg at the winter solstice.

Therefore, increasingly towards the north, the azimuthal positions of the sunrise are so close to the true north or south at the summer and winter solstices, respectively, that it is often hard to distinguish deliberate Cardinal orientations from solar ones. In some cases they are, in effect, the same with the precision with which the orientations can be measured due to the preservation of the structures. This is especially true for the winter solstice orientations, if the horizon is elevated.

An interesting point can be raised based on this movement of the sun close to the horizon line. It is known from the Greek historian Diodorus Siculus that the movement of the moon close to the horizon was considered significant by the builders of at least one Megalithic temple, identified by A. Burl as the ring of Callanish on Lewis, Outer Hebrides.[4] Finland is too far north for the extreme points of the moon to be used for orientating buildings, but a similar effect to the one described by Diodorus Siculus is produced by the movements of the sun, which can be used for orientation purposes. One reason for building these monuments in Ostrobothnia could be related to this kind of movement of the sun at these latitudes.

The society that built the Giant's Churches was Neolithic, but agriculture had already arrived in the region. The first signs of grain cultivation in the Northern Finland are from Puolanka in 2200 BCE.[5] This place is about 70 km from the Giant's Churches in Tyrnävä and Muhos. The Northern Mesolithic society probably cooperated with the first Neolithic farmers by trade, which lead to the adaptation of new ideas, which eventually altered the social structure and cultural practices, including the concept of time.[6] At the time, social complexity increased. The building of monumental structures may reflect this change.[7] The possible cultural transformation of the hunter-gatherer society by obtaining, amongst other influences, calendrical information from the Neolithic farming society could explain the building of large structures with solar orientations. Thus, the Giant's Churhes of Ostrobothnia could ultimately be seen as part of the early Neolithic enclosure-building, which first manifested itself in the rondels of Northern Germany in 5000-4500, and continued in the later enclosures in the region and in the constructions of the Megalithic culture in Western Europe.[8] However, it is too early to say whether the Giant's Churches were used primarily as cult places. They may have had other functions, and orientating the structures towards important solar dates may have been a "mode of fashion" rather than the manifestation of new religious beliefs. Further research on the sites is required to understand the meaning of the solar orientations of the Giant's Churches observed in this study.

*References*

Baldia, M.: A spatial analysis of megalithic tombs; Ph.D. Dissertation, Southern Methodist University, Dallas (1995).

Bertemes, F., Biehl, P. F., Northe, A., and Schröder, O.: Die neolithische Kreisgrabenanlage von Goseck, Landkreis Weißenfels; Archäologie in Sachsen-Anhalt 2/2004 (2004), 137-145.

Bertemes, F., and Spatzier, A.: Pömmelte – ein mitteldeutsches Henge-Monument aus Holz; Archäologie in Deutschland 6/2008 (2008), 6-11.

Burl, A.: A guide to the stone circles of Britain, Ireland and Brittany; Yale University Press, London (1995).

Europaeus, A.: Paavolan pitäjän "jättiläiskirkot"; Suomen Museo XX (1913).

---

[4] Burl 1995, p. 150.
[5] Vuorela 2002, p. 84-87.
[6] Okkonen 2003, p. 219-226.
[7] Okkonen 2003, p. 223.
[8] See, e.g.: Baldia 1995; Burl 1995; Hoskin 2001; Bertemes et al. 2004; Ruggles 2005; Bertemes & Spatzier 2008; Pasztor & Barna 2008.

Hoskin, M.: Tombs, temples and their orientations; Ocarina Books, London (2001).

Okkonen, J.: Jättiläisen hautoja ja hirveitä kiviröykkiöitä – Pohjanmaan muinaisten kivirakennelmien arkeologiaa; väitös; Oulu (2003).

Pásztor, E., and Barna, J. P.: Orientation of the circular enclosures of the late Neolith Lengyel culture in the Carpathian Basin; Antiquity 82 (2008), 910-924.

Ruggles, C. L. N.: Ancient Astronomy: An Encyclopedia of Cosmologies and Myth; ABC-CLIO, London (2005).

Vuorela, I: Luonnon kerrostumat säilövät menneisyyttä. Ennen, muinoin. Miten menneisyyttämme tutkitaan. Ed. R. Grünthal, Helsinki (2002).

| # | Name | Location | Shape | Size (m) | Long axis (deg) | Horizon | Solar event | Ori. of axis | Other ori. (deg) | Horizon | Solar event | Name of event |
|---|------|----------|-------|----------|-----------------|---------|-------------|--------------|------------------|---------|-------------|---------------|
| 1 | Rajakangas | Haukipudas | Q2 | 32x26 | 50 | | x+1 | Vappu srise | 50 | | x+1 | Vappu srise |
| | | | | | | | | | 268 | 1 | x+1 | Eqx sset |
| 2 | Kastelli | Pattijoki, Raahe | Q | 62x36 | 13 | | x | SS srise | 3 | | x-9 | SS srise |
| | | | | | | | | | 42 | | x-6 | Vappu srise |
| | | | | | | | | | 103 | 1 | x+12 | Eqx srise |
| | | | | | | | | | 169 | 1 | x | WS srise |
| | | | | | | | | | 193 | | x-4 | WS sset |
| 3 | Kiviojankangas | Pattijoki, Raahe | Open R | 40x21 | 153 | | x-4 | WS srise | o30 | | x+11 | SS srise |
| | | | | | | | | | 65 | | x+14 | Vappu srise |
| | | | | | | | | | o134 | | x+1 | Kekri srise |
| | | | | | | | | | 147 | | x-10 | WS srise |
| 4 | Pesuankangas | Ruukki, Siikajoki | Q | 33x24 | 138 | | x+5 | Kekri srise | 133 | | x | Kekri srise |
| | | | | | | | | | 313 | | x | Vappu sset |
| 5 | Linnakangas* | Ruukki, Siikajoki | Egg-shaped | 27x18 | 32 | | x+16 | SS srise | 85 | | x-3 | Eqx srise |
| | | | | | | | | | 300? | | x-10 | Vappu sset |
| 6 | Kettukangas* | Raahe | Q | 30x20 | 45 | | x-3 | Vappu srise | s201 | | x | WS sset |
| | | | | | | | | | 190 | | x-8 | WS sset |
| 7 | Pikku Jakenaro | Raahe | Q2 | 33x16 | 175 | | x+12 | WS srise | 175 | | x+13 | WS srise |
| | | | | | | | | | 350 | | x+2 | SS sset |
| | | | | | | | | | 257 | | x-13 | Eqx sset |
| 8 | Linnamaa* | Temmes, Liminka | Q2 | 40x28 | 127 | | x-3 | Kekri srise | 307 | | x-3 | Vappu sset |
| | | | | | | | | | 127? | | x-3 | Kekri srise |
| 9 | Mustosenkangas | Liminka | Q2 | 37x23 | 138 | | x+5 | Kekri srise | 50 | | x | Vappu srise |
| | | | | | | | | | 230 | | x | Kekri sset |
| | | | | | | | | | 138 | | x+5 | Kekri srise |
| | | | | | | | | | 318 | | x+5 | Vappu sset |
| 10 | Metelinkangas | Tyrnävä | Q | 40x22 | 72 | | x-16 | Eqx srise | 238 | | x+8 | Kekri sset |
| | | | | | | | | | 145 | 2 | x+6 | Kekri srise |
| 11 | Hangaskangas | Kannus | Open R | 35x30 | 125 | | x-7 | Kekri srise | so112 | | x-17 | Kekri srise |
| | | | | | | | | | so175 | | x+16 | WS srise |
| | | | | | | | | | 122 | | x-10 | Kekri srise |
| | | | | | | | | | 138 | | x+6 | Kekri srise |
| 12 | Hiidenlinna | Himanka | Q | 45x30 | 78 | | x-11 | Eqx srise | 89 | | x | Eqx srise |
| | | | | | | | | | 245 | | x+14 | Kekri sset |
| | | | | | | | | | 163? | | x+3 | WS srise |
| | | | | | | | | | 180 | | x+20 | WS srise |
| | | | | | | | | | 341 | | x | SS sset |
| | | | | | | | | | s51 | | x | Vappu srise |
| | | | | | | | | | s114 | | x-15 | Kekri srise |
| 13 | Honkobackaharju N | Kruunupyy | Q | 17x10 | 166 | | x+8 | WS srise | 166 | | x+8 | WS srise |
| | | | | | | | | | 346 | | x+5 | SS sset |
| 14 | Brantbacken 1 | Kruunupyy | Q | 13x10 | 171 | | x+11 | WS srise | 168 | | x+10 | WS srise |
| | | | | | | | | | 353 | | x+12 | SS sset |
| 15 | Brantbacken 2 | Kruunupyy | Q | 12x10 | 54 | | x+1 | Vappu srise | 54 | | x+1 | Vappu srise |
| 16 | Ollisbacken 1 | Kruunupyy | Q | 30x20 | 49 | | x | Vappu srise | 49 | | x+1 | Vappu srise |
| | | | | | | | | | 229 | | x | Kekri sset |
| 17 | Ollisbacken 2 | Kruunupyy | Q | 30x20 | 54 | | x+1 | Vappu srise | 54 | | x+1 | Vappu srise |
| | | | | | | | | | 239 | | x+7 | Kekri sset |
| 18 | Ollisbacken 3 | Kruunupyy | Q | 30x20 | 53 | | x | Vappu srise | 53 | | x | Vappu srise |
| 19 | Snårbacken | Kruunupyy | Q | 17x10 | 116 | 1 | x-16 | Kekri srise | 281 | | x+9 | Eqx sset |
| 20 | Högryggen | Kruunupyy | Q2 | 36x20 | 172 | | x+12 | WS srise | 174 | | x+14 | WS srise |
| | | | | | | | | | 354 | | x+13 | SS sset |
| 21 | Svedjebacken | Pedersöre | Q2 | 58x34 | 111 | | x-12 | Kekri srise | 111 | | x-17 | Kekri srise |
| | | | | | | | | | 291 | | x-17 | Vappu sset |
| | | | | | | | | | 206 | | x | WS sset |
| | | | | | | | | | s26? | | x | SS srise |
| 22 | Jäknabacken | Pedersöre | Q | 65x35 | 148 | | x-4 | WS srise | 1 | | x-19 | SS srise |
| | | | | | | | | | 148 | | x-4 | WS srise |
| 23 | Storbacken 1 | Evijärvi | Q | 30x20 | 64 | | x+11 | Vappu srise | | | | |

**Table 1.** Orientations of the Giant's Churches. The azimuths of the orientations are given in full degrees. Abbreviations: WS = winter solstice; SS = summer solstice; srise = sunrise; sset = sunset. The azimuth of the closest solar event is given as x+n = m, where x is the event given in 'Name of Event', and n is the difference between this event and the measured orientation m. 'o' denotes an orientation towards an open wall, and 's' denotes an orientation towards a "sacristy".

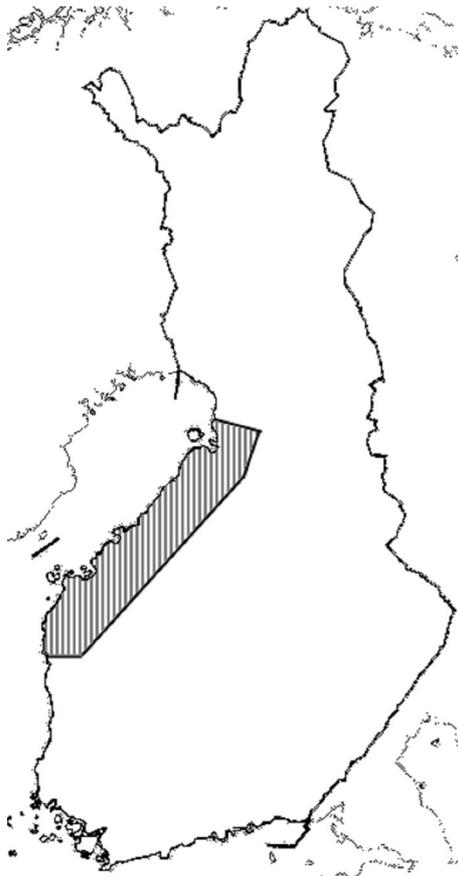

**Figure 1.** Area of Finland where Giant's Churches are found.

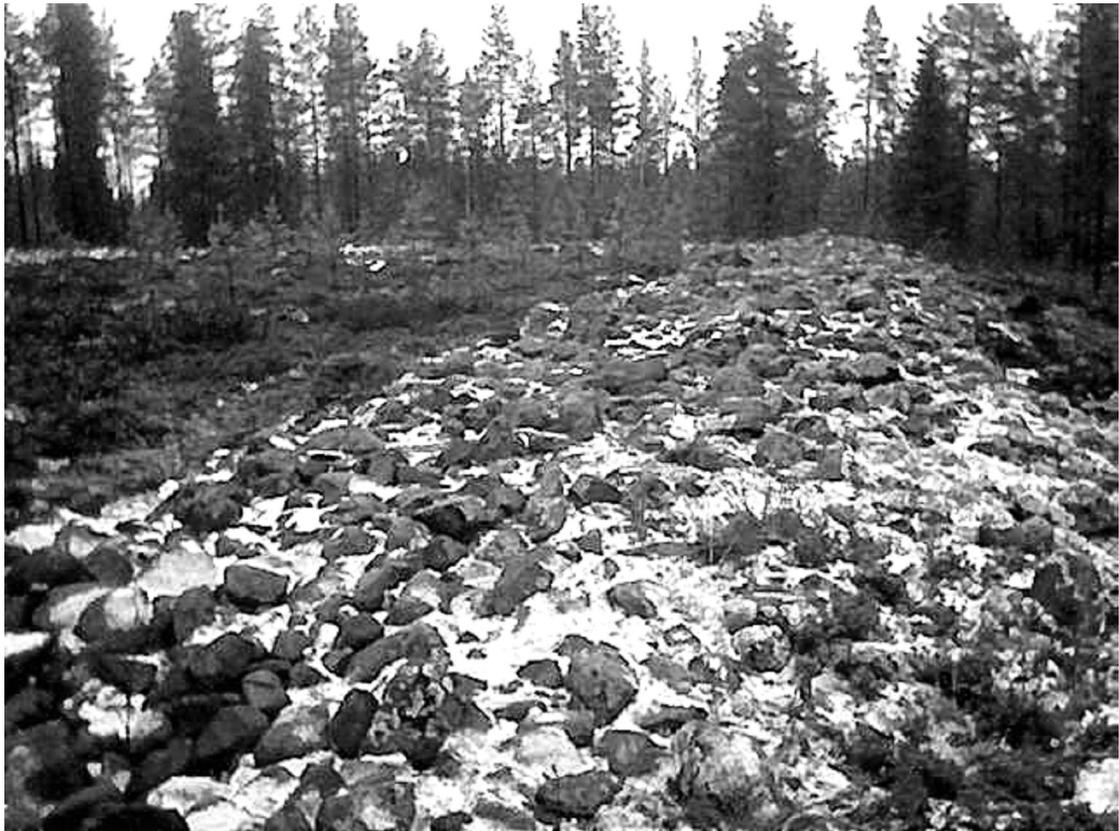

**Figure 2.** Stone walls of the Giant's Church of Kastelli in Raahe. The greatest length of the structure is 62 m.

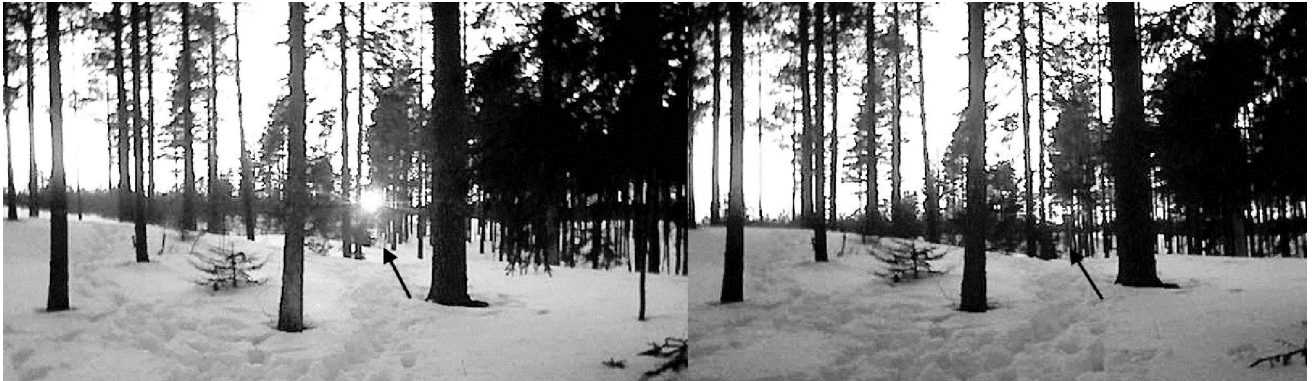

**Figure 3.** Sun setting behind the western gate (az 268 deg) of the Giant's Church of Rajakangas in Haukipudas at the spring equinox in 2009. The arrow points to a rod placed in the snow to mark the position of the gate.

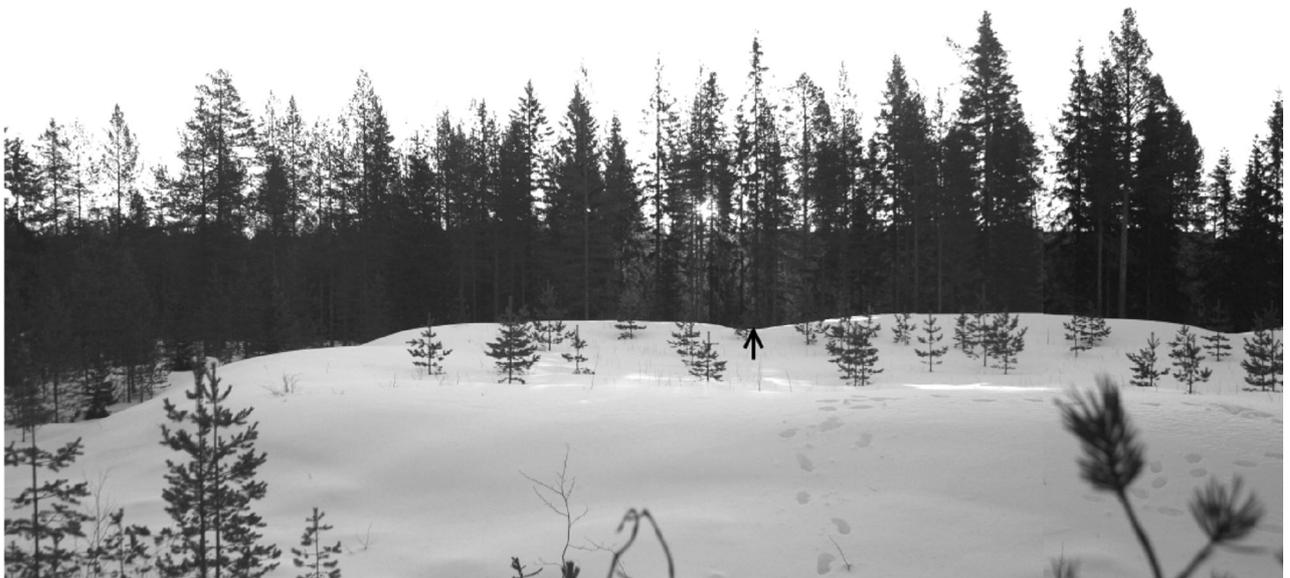

**Figure 4.** Sun rising behind the eastern gate of Kastelli in Raahe at the spring equinox. The orientation of the gate (az 103 deg), marked with an arrow in the figure, is towards the sunrise 11 days before the spring equinox as measured from the centre of the structure. The orientation may be related to the calculation of the solar year using lunar months.